\documentclass[12pt]{article}
\usepackage[utf8]{inputenc}
\usepackage[T1]{fontenc}
\usepackage{lmodern}
\usepackage{amsmath,amssymb,amsthm}
\usepackage[authoryear,round]{natbib}
\usepackage[colorlinks=true,citecolor=blue,linkcolor=blue,urlcolor=blue]{hyperref}
\usepackage[margin=25mm]{geometry}
\usepackage{setspace}
\newtheorem{assumption}{Assumption}
\newtheorem{theorem}{Theorem}
\newtheorem{lemma}{Lemma}
\theoremstyle{definition}
\newtheorem{definition}{Definition}
\newtheorem{remark}{Remark}

\newcommand{\R}{\mathbb R}

\title{Collateral Equilibrium with a Continuum of States\\ and No Ex Post Survival Assumption}
\author{Jaime Orrillo\thanks{The author has been a summer visiting professor at IMPA every year since 2000, except during the COVID-19 pandemic. E-mail: \texttt{orrillo@p.ucb.br}, \texttt{jjoc@impa.br}.}\\[2pt] \normalsize Catholic University of Bras\'ilia}
\date{\today}

\begin{document}
\maketitle

\begin{abstract}
Existence results for incomplete markets with a continuum of states require ex post incomes to be nonnegative, or endowments to be bounded away from zero in every state. Such conditions rule out disaster states, in which some agents receive no income at all. We show that, with a single good and collateralized securities in the sense of Geanakoplos and Zame, an equilibrium exists when second-period endowments are arbitrary nonnegative functions of the state: they may vanish on any set of states, including all of them. Only first-period endowments are required to be positive; no survival assumption is imposed ex post. The reason is that non-recourse collateral makes every portfolio solvent in every state, and with one good this is all the limit argument needs.

\end{abstract}

\section{Introduction}

In his account of the mortgage market, \citet{Geanakoplos1997} identifies states of the world with the scenarios that investors compute for future cash flows, and observes that ``there is an infinity of potential scenarios''; any finite list of them gives, at best, anecdotal information about future returns. A continuum of states is therefore not a technical generalization but the natural model of the uncertainty faced by collateralized markets. Asset pricing works with it routinely: lognormal returns, diffusions, rare disasters \citep{Rietz1988,Barro2006}. In disaster states the income of some agents is wiped out. General equilibrium theory with incomplete markets and a continuum of states, however, rests on assumptions that exclude such states. \citet{MasColellZame1996} and \citet{Monteiro1996} require ex post endowments to be nonnegative for the portfolios agents trade; \citet{AraujoMonteiroPascoa1997} drop this requirement by allowing default subject to utility penalties, as in \citet{DubeyGeanakoplosShubik2005}; and \citet{Orrillo2001}, who replaces it by collateral, assumes that second-period endowments are bounded away from zero, uniformly across states; \citet{Orrillo2009}, who extends this approach to an infinite horizon, keeps the same uniform lower bound on endowments. This assumption is strong and exclusive: it rules out economies where some agent may receive nothing in some states.

This note removes it. We work in the collateral framework of \citet{GeanakoplosZame2014} with a single good, the standard setting of financial economics, and a continuum of states. Securities are non-recourse: deliveries are the minimum of the promise and the value of the collateral. We show that an equilibrium exists when second-period endowments are arbitrary nonnegative, bounded, measurable functions of the state. They may vanish on any set of states. Only first-period endowments must be positive.

The argument is simple, and the simplicity is the point. With non-recourse collateral, a borrower never delivers more than the collateral he holds, so ex post income is at least the endowment for every portfolio, in every state: every portfolio is solvent, as already observed by \citet{Orrillo2009}, and the positivity of ex post income that \citet{MasColellZame1996} must assume holds by construction. With one good, consumption in each state equals ex post income, which is determined by a finite-dimensional portfolio. The limit of truncated economies then needs neither Lagrange multipliers nor Fatou's lemma, which is where a lower bound on endowments was used in \citet{Orrillo2001}; it needs only Lebesgue's differentiation theorem and dominated convergence.

For finitely many states, \citet{GottardiHens1996} study existence without survival assumptions through quasi-equilibria and irreducibility. For collateral economies with finitely many states, \citet{Orrillo2011} shows that, when every agent has a nonzero date-0 endowment, neither a strictly positive social endowment nor second-period endowments are needed. This note extends that result to a continuum of states, with a single good. With one good, a positive date-0 endowment is the survival assumption at date 0, and we keep it; what the present note removes is survival ex post.

\section{The economy}

There are two dates. Uncertainty at date 1 is described by the Lebesgue measure space $(S,\mathcal S,\lambda)$, $S=[0,1]$. There is a single good at each date. The good is durable: one unit held at date 0 becomes $y(s)\ge0$ units in state $s$. When $y\equiv0$ the good is perishable; when $y>0$ it may be read as housing, a stored commodity or a productive asset.

There are $H$ agents. Agent $h$ has endowments $e_0^h\in\R_+$ at date 0 and $e^h:S\to\R_+$ at date 1, and utility
\[
U^h(x_0,x)=u_0^h(x_0)+\int_Su^h\big(s,x(s)\big)\,\lambda(ds).
\]

There are $J$ securities. Security $j$ promises $A_j(s)\ge0$ units of the good in state $s$ and requires collateral $c_j>0$: a seller of $\varphi_j$ units must hold $c_j\varphi_j$ units of the good at date 0. As in \citet{GeanakoplosZame2014}, the collateral is the only means of enforcement, so a seller delivers
\[
D_j(s)=\min\{A_j(s),\,c_j\,y(s)\}
\]
per unit in state $s$. The good is the numeraire at every date and $q\in\R_+^J$ are security prices. Agent $h$ chooses $x_0\ge0$, purchases $\theta\in\R_+^J$, sales $\varphi\in\R_+^J$ and date-1 consumption $x\ge0$, subject to
\begin{gather}
x_0+q\cdot\theta\le e_0^h+q\cdot\varphi,\qquad \sum_jc_j\varphi_j\le x_0, \label{eq:b0}\\
x(s)\le I^h(s):=e^h(s)+y(s)\,x_0+\sum_jD_j(s)(\theta_j-\varphi_j)\qquad\text{a.e.} \label{eq:b1}
\end{gather}
The first line is the date-0 budget and the collateral constraint of \citet{GeanakoplosZame2014}; the second is the budget in state $s$, where $I^h(s)$ is ex post income.

\begin{lemma}[Solvency]\label{lem:solv}
For every plan satisfying \eqref{eq:b0}, $I^h(s)\ge e^h(s)\ge0$ for every $s$.
\end{lemma}

\begin{proof}
Since $D_j(s)\le c_jy(s)$ and $\theta\ge0$, $I^h(s)\ge e^h(s)+y(s)\big(x_0-\sum_jc_j\varphi_j\big)\ge e^h(s)$ by the collateral constraint.
\end{proof}

\begin{definition}
An equilibrium is a price vector $q\in\R_+^J$ and plans $(x_0^h,\theta^h,\varphi^h,x^h)_h$ such that each plan maximizes $U^h$ subject to \eqref{eq:b0}--\eqref{eq:b1}, and
\[
\sum_hx_0^h=\sum_he_0^h,\qquad \sum_h\theta^h=\sum_h\varphi^h,\qquad \sum_hx^h(s)=\sum_he^h(s)+y(s)\sum_he_0^h\quad\text{a.e.}
\]
\end{definition}

\begin{assumption}\label{as}
Throughout the paper we maintain the following assumptions.
\begin{enumerate}
\item[(A1)] For each $h$: $u_0^h:\R_+\to\R$ is continuous, concave and strictly increasing; $u^h(s,\cdot):\R_+\to\R$ is continuous, concave and strictly increasing for every $s$; $u^h$ is a Carath\'eodory function, measurable in $s$ and continuous in $v$; and for every $v\ge0$, $s\mapsto u^h(s,v)$ is bounded.
\item[(A2)] For each $h$: $e_0^h>0$, and $e^h:S\to\R_+$ is measurable and essentially bounded.
\item[(A3)] $y:S\to\R_+$ and $A_j:S\to\R_+$ are measurable and essentially bounded, and $c_j>0$ for each $j$.
\end{enumerate}
\end{assumption}

No lower bound is imposed on second-period endowments: $e^h$ may vanish on any measurable set, including $S$ itself.

\begin{theorem}\label{thm}
Under (A1)--(A3), an equilibrium exists.
\end{theorem}

\section{Proof}

Write $E_0=\sum_he_0^h$, $\bar e=\operatorname{ess\,sup}\sum_he^h$, $\bar y=\operatorname{ess\,sup}y$ and $B=\bar e+\bar yE_0+H$.

\emph{Step 1: truncated economies.} For $N\ge1$ let $I_n=[(n-1)/N,n/N)$, $n=1,\dots,N$. The economy $\mathcal E^N$ has $N$ states of probability $1/N$; in state $n$ the data are the averages $e_n^{hN}=N\int_{I_n}e^h\,d\lambda+1/N$, $y_n^N=N\int_{I_n}y\,d\lambda$ and $A_{jn}^N=N\int_{I_n}A_j\,d\lambda$, with deliveries $D_{jn}^N=\min\{A_{jn}^N,c_jy_n^N\}$, and agent $h$ has utility $u_0^h(x_0)+\sum_n\int_{I_n}u^h(s,x_n)\,\lambda(ds)$. The term $1/N$ makes aggregate endowments strictly positive in every state. $\mathcal E^N$ is a collateral economy with finitely many states that satisfies the assumptions of \citet[Theorem 1]{GeanakoplosZame2014}; utilities in $\mathcal E^N$ are continuous, concave and strictly increasing by (A1) and dominated convergence. In particular, the endowment of agent $h$ in $\mathcal E^N$ is $(e_0^h,e_1^{hN},\dots,e_N^{hN})\in\R_+^{1+N}\setminus\{0\}$, with $e_0^h>0$. By that theorem, $\mathcal E^N$ has an equilibrium. Let $(q^N,(x_0^{hN},\theta^{hN},\varphi^{hN},x_n^{hN})_{h,n})$ be an equilibrium of $\mathcal E^N$. Since $u^h(s,\cdot)$ is strictly increasing, $x_n^{hN}$ equals ex post income $I_n^{hN}$.

\emph{Step 2: bounds.} Market clearing at date 0 gives $x_0^{hN}\le E_0$; the collateral constraint gives $\varphi_j^{hN}\le E_0/c_j$; market clearing for securities gives $\theta_j^{hN}\le HE_0/c_j$. Moreover $q_j^N\le c_j$: if $q_j^N>c_j$, an agent could sell one more unit of $j$ and raise date-0 consumption by $q_j^N>c_j$, of which $c_j$ units serve as collateral; his ex post income does not fall in any state, because the extra good returns $y_n^Nq_j^N\ge c_jy_n^N\ge D_{jn}^N$. This contradicts optimality. Finally $0\le x_n^{hN}\le B$ for $N\ge1$.

\emph{Step 3: limits.} Along a subsequence, $q^N\to q$, $x_0^{hN}\to x_0^h$, $\theta^{hN}\to\theta^h$ and $\varphi^{hN}\to\varphi^h$. Let $e^{hN}(s)$, $y^N(s)$, $A_j^N(s)$ be the step functions equal to the averages on each $I_n$ (without the term $1/N$ for $e$). By Lebesgue's differentiation theorem, which applies because the interval $I_n$ that contains $s$ lies in $(s-1/N,s+1/N)$ and has half its measure, they converge to $e^h(s)$, $y(s)$, $A_j(s)$ for a.e.\ $s$, and therefore $D_j^N(s)\to D_j(s)$, since $\min$ is continuous. Hence the truncated consumption $x^{hN}(s)$, the step function equal to $x_n^{hN}$ on $I_n$, converges for a.e.\ $s$ to
\[
x^h(s)=e^h(s)+y(s)\,x_0^h+\sum_jD_j(s)(\theta_j^h-\varphi_j^h)=I^h(s).
\]
Passing to the limit in the budget and market-clearing conditions of $\mathcal E^N$ shows that $(x_0^h,\theta^h,\varphi^h,x^h)$ satisfies \eqref{eq:b0}--\eqref{eq:b1} at $q$, and that all markets clear.

\emph{Step 4: optimality.} For a plan $\pi=(x_0,\theta,\varphi)$ satisfying \eqref{eq:b0} at prices $q'$, let $V^N(\pi)=u_0^h(x_0)+\int_Su^h(s,I^{hN}_\pi(s))\,\lambda(ds)$, where $I^{hN}_\pi$ is the step function of ex post incomes in $\mathcal E^N$, and define $V(\pi)$ in the same way with $I^h_\pi$. Incomes lie in $[0,B']$ for some $B'$ that depends only on the bounds of Step 2 and on the plan, and by (A1) and monotonicity $|u^h(s,c)|\le\max\{|u^h(s,0)|,|u^h(s,B')|\}$, a bounded function of $s$. By Step 3 and dominated convergence, $V^N(\pi^{hN})\to V(\pi^h)$, where $\pi^{hN}$ and $\pi^h$ are the equilibrium plans of agent $h$ in $\mathcal E^N$ and their limit.

Let $\pi'=(x_0',\theta',\varphi')$ satisfy \eqref{eq:b0} at $q$, and for $\eta\in(0,1)$ let $\pi'_\eta=(1-\eta)\pi'$. Then $\pi'_\eta$ satisfies the collateral constraint, and
\[
(1-\eta)\big(x_0'+q^N\cdot\theta'-q^N\cdot\varphi'\big)\le(1-\eta)e_0^h+(1-\eta)\,|q^N-q|\,|\theta'-\varphi'|<e_0^h
\]
for $N$ large, because $e_0^h>0$. Hence $\pi'_\eta$ is feasible at date 0 in $\mathcal E^N$, and by Lemma~\ref{lem:solv} it is solvent in every state. Optimality in $\mathcal E^N$ gives $V^N(\pi^{hN})\ge V^N(\pi'_\eta)$. By dominated convergence $V^N(\pi'_\eta)\to V(\pi'_\eta)$, so $V(\pi^h)\ge V(\pi'_\eta)$. Letting $\eta\to0$, continuity of $u_0^h$ and dominated convergence give $V(\pi^h)\ge V(\pi')$. Since $u^h(s,\cdot)$ is increasing, consuming ex post income is optimal given the portfolio. Hence $\pi^h$, with consumption $x^h$, is optimal at $q$, and the limit is an equilibrium. \qed

\section{Remarks}

\begin{remark}[What collateral does]
Lemma~\ref{lem:solv} is the whole contribution of collateral. Without it, a seller of a security that promises more than his income in some states could not honor the promise, and portfolios would have to be restricted state by state, as in \citet{MasColellZame1996}. With non-recourse collateral, solvency holds for every portfolio, and the positivity of ex post income becomes a consequence of the market structure rather than an assumption on endowments.
\end{remark}

\begin{remark}[Disaster states]
The theorem allows $e^h=0$ on sets of positive measure, and even $\sum_he^h=0$ there. In such states consumption is financed only by the durable good and by collateralized claims: this is the content of rare-disaster models \citep{Rietz1988,Barro2006}, now in general equilibrium with default. Equilibrium consumption may be zero in some states; nothing in the argument requires it to be positive.
\end{remark}

\begin{remark}[First-period endowments]
Positivity of $e_0^h$ is used twice: in Step 1, where Theorem~1 of \citet{GeanakoplosZame2014} is applied to $\mathcal E^N$, and in Step 4, to make scaled deviations strictly feasible at date 0. With one good it is the survival assumption at date 0, and it cannot be dropped. If some agent has $e_0^h=0$, his date-0 budget set may fail to be lower hemicontinuous in security prices at $q_j=c_j$: below this price he cannot sell security $j$, at it he can. This is the classical survival problem, treated with quasi-equilibria by \citet{GottardiHens1996}. With $L>1$ goods, the corresponding condition $e_0^h\neq0$ is weaker than survival at date 0.
\end{remark}

\begin{remark}[Recourse loans]
The theorem relies on loans being non-recourse: a borrower never delivers more than the value of his collateral, which is what Lemma~\ref{lem:solv} uses. When lenders have recourse beyond the collateral, deliveries may exceed collateral returns, ex post income may fall below the endowment, and solvency is no longer automatic. Recourse must then be enforced by penalties, as in \citet{DubeyGeanakoplosShubik2005}. \citet{PascoaSeghir2009} show that, with collateral and harsh utility penalties for default, borrowers may pay more than the value of the collateral, so that loans can exceed collateral costs and Ponzi schemes become possible. Whether an equilibrium exists with recourse loans, a continuum of states and second-period endowments that may vanish is a different question, which the present argument does not answer.
\end{remark}

\begin{remark}[Many goods]
With $L>1$ goods in each state, consumption is no longer determined by the portfolio alone; spot prices and the choice of bundles must also converge, and the argument requires Lagrange multipliers. A lower bound on endowments then reappears in the bound on multipliers \citep{Orrillo2001,Orrillo2009}. Removing it in that case is left open.
\end{remark}

\end{document}